# Harnessing the Potential of Blockchain in DevOps: A Framework for Distributed Integration and Development

**Muhammad Shoaib Farooq[1], Usman Ali[1]**
[1]Department of Artificial Intelligence, School of System and Technology, University of Management and Technology, Lahore, 54000, Pakistan
Corresponding author: Muhammad Shoaib Farooq (shoaib.farooq@umt.edu.pk)

**ABSTRACT** As the use of DevOps practices continues to grow, organizations are seeking ways to improve collaboration, speed up development cycles, and increase security, transparency, and traceability. Blockchain technology has the potential to support these goals by providing a secure, decentralized platform for distributed integration and development. In this paper, we propose a framework for distributed DevOps that utilizes the benefits of blockchain technology that can eliminate the shortcomings of DevOps. We demonstrate the feasibility and potential benefits of the proposed framework that involves developing and deploying applications in a distributed environment. We present a benchmark result demonstrating the effectiveness of our framework in a real-world scenario, highlighting its ability to improve collaboration, reduce costs, and enhance the security of the DevOps pipeline. Conclusively, our research contributes to the growing body of literature on the intersection of blockchain and DevOps, providing a practical framework for organizations looking to leverage blockchain technology to improve their development processes.

## I. Introduction

Software development has been becoming more complex day by day, and overcoming the complexities and challenges of software development, one of the best approaches is using DevOps. DevOps is a relatively new technique that is used to give new software updates with maximum reliability as well as accuracy while doing continuous integration and continuous development (CI/CD) [1]. DevOps combines the Development Team and Operation Team's people, processes, and technology to work in a collaborative environment while giving value to customers [2]. The DevOps eliminate the problem of clashes between the Dev Team and Op Team, and because of that, the project can be completed successfully. On the other hand, Distributed DevOps is an approach to developing and operating software systems that involve distributed teams working together to build, deploy, and maintain the software in a collaborative manner [2]. In traditional DevOps models, developers and operations teams work together to build and deploy software, but this collaboration often occurs within a single organization. In contrast, distributed DevOps involves teams that are geographically dispersed and may not be part of the same organization. The potential advantage of distributed DevOps is that it allows organizations to tap into a wider pool of talent and expertise, as they are not limited to a single location. It can also enable organizations to be more agile and responsive to changing market conditions, as they can quickly bring new resources online as needed.

The adoption of DevOps practices in distributed teams can bring numerous benefits, but it also presents significant challenges around transparency, trust, security, and traceability. To ensure the success of distributed DevOps, it is critical to establish trust and transparency, particularly when teams are working across different locations and time zones. Security is also of paramount importance to prevent any malicious actors from disrupting or compromising the project. Moreover, traceability is essential for ensuring efficient product recalls and enabling all stakeholders to work in a more trustworthy environment.

The development process is susceptible to cyber threats and malicious attacks, which can result in data breaches, system downtime, and financial losses if there is no security in the system. Moreover, without appropriate traceability, it can be difficult to identify the root cause of errors or bugs, which can lead to delays in the development cycle and increased costs. The lack of transparency and trust can result in poor collaboration and communication between team members, leading to misunderstandings and misaligned goals. This can also result in delays, wasted effort, and decreased productivity.

Blockchain is a distributed ledger that provides a secure and transparent platform for recording and verifying transactions [4]. It enables developers to create tamper-proof records of all changes made to code, providing a secure and immutable audit trail that can be traced back to the original source.

We have proposed framework that combines DevOps with blockchain that can utilize the benefits of blockchain technology to create a more secure, transparent, and efficient development environment. This framework involves developing and deploying applications in a distributed environment, leveraging blockchain's decentralized and tamper-proof properties to create a trustworthy and traceable audit trail. The framework incorporates smart contracts, which can automate the testing and deployment of code, reducing the time and effort required to complete the development cycle. It also includes consensus mechanisms, which enable multiple parties to agree on the state of the code and ensure that all parties have access to the same information. The framework includes robust security measures, such as cryptographic hashing and encryption, to protect the code and sensitive data from cyber threats and malicious attacks.

The novelty of our proposed framework is that it simplifies communication and data sharing for distributed teams in a secure, traceable, and immutable environment. We offer a strategy for implementing DevOps with Blockchain, reducing conflicts between clients and software development teams. Our solution addresses traceability, trust, and transparency issues to create a better work environment.

In the remaining paper, the sections are as follows: In Section 2, the related work has been presented, and in Section 3, we have elaborated Preliminaries used in the model. The proposed framework is shown in Section 4, and Section 5 contains the implementation and performance. Furthermore, in Section 6, the discussion is shown, and finally, Section 7 describes the conclusion and future work.

**II. Related Work**

There has been a significant amount of research done on the intersection of blockchain and software development models. Here are a few examples of research in this area:

Muhammad Azeem Akbar et. al. [6] have presented a paper in which the study investigates the potential benefits of using blockchain technology in a DevOps paradigm. A framework is proposed which helps merge the characteristics of blockchain with the DevOps paradigm. This provides an effective means for the adoption of blockchain in DevOps while ensuring its advantages over other solutions are maximized. The research paper can be a good start, however, the proposed framework lacks details so it was not clear how the framework is going to work.

Sandip Bankar and Deven Shah [7] also presented a paper that intersects blockchain with DevOps. In this research paper, it is described that blockchain technology can be used in DevOps for software development. The benefits of this include improved quality and performance, as well as increased security. This is achieved by storing all project artifacts in a decentralized and secure blockchain environment. DevOps and blockchain together can form a highly advanced business solution that can revolutionize services, operations, and collaborations. Although, it is not suggested for distributed DevOps.

Maximilian Wöhrer and Uwe Zdun [8] investigated current practices and solution approaches for an efficient blockchain-oriented DevOps procedure by combining gray literature and DevOps application studies from pertinent GitHub projects. Their research has shown that core DevOps concepts and activities are similar to other areas, but more rigorous testing and differentiated deployment practices are required due to the inherent immutability of the blockchain. Overall, the study suggests that a structured DevOps approach is feasible with already established CI/CD solutions that orchestrate the right tools. But the findings may not be applicable to all organizations and their specific needs and requirements for implementing a blockchain-oriented DevOps procedure.

Md Jobair Hossain Faruk, Santhiya Subramanian, Hossain Shahriar et. al. [9] examine the impact of existing software engineering processes. They consider the need to adopt new concepts and evolve current software engineering processes for blockchain systems. The study also looks at the role of software project management in the development of blockchain-oriented software. The findings suggest that using state-of-the-art techniques in software processes for futuristic technologies may be challenging, and more research is needed to improve software engineering processes and methodologies for these novel technologies. However, this was a systematic study and does not present any kind of framework.

These are just a few examples of the many research efforts focused on blockchain and software development models. It is worth noting that while these studies have the potential to bring many benefits to the software development process, they do not present any framework that is for Distributed DevOps where teams are not in the same place or even time zone.

The novelty of our proposed framework is that it could make it simpler for distributed teams to communicate and share data in a secure, traceable, and immutable environment. We have presented a comprehensive strategy for implementing DevOps methods with respect to Blockchain to successfully execute projects while reducing conflicts between Clients and Software Development Teams. We have solved the issues of traceability, trust, and transparency of DevOps so that everyone can work in a trusted and more secure environment. DevOps phases; project initiation, continuous development, continuous integration & delivery, continuous deployment, and continuous monitoring are shown and how they will work with the Blockchain is also described. Using smart contracts, we have also resolved payment-related issues. In addition, IPFS is used to address the scalability issue of the Blockchain.

### III. Preliminaries
This section presents the preliminaries for the blockchain-based framework. The major components that will be used in this framework will be described in this section including, IPFS, Blockchain, Smart Contracts, and Decentralized Applications.

#### A. IPFS (InterPlanetary File System)
IPFS (InterPlanetary File System) is a platform that enables decentralized storage and sharing of information [10]. It works by creating a distributed system where any device can store files, and users can access these files from anywhere in the world. IPFS uses a peer-to-peer network to make it easy for people to share files without having to go through central authorities or servers. This makes it resistant to censorship and seizure, as well as faster than traditional file systems because it does not need to wait for requests to be processed.

#### B. Blockchain and Smart Contracts
Blockchain is a decentralized ledger system that offers safe and transparent data storage as well as the transfer of assets and other types of information [11]. Through the use of cryptographic techniques, it produces a record of transactions that cannot be altered. This record takes the form of a chain of blocks, each of which references the block that came before it.

Smart contracts are agreements that may execute themselves and have their conditions specified in code [12]. They are kept on a blockchain. They make it possible to execute complicated transactions in a way that is tamper-proof, transparent, and efficient, which could potentially reduce the need for intermediaries.

#### C. Decentralized Applications
Decentralized Applications also known as DApps are the applications that are presented on the distributed network. These applications are made with smart contracts and they also have a front and backend that also run on a decentralized peer-to-peer network.

#### D. Jenkins
Jenkins is a widely-used open-source automation tool for CI/CD in software development. Its flexibility, extensibility, and support for pipeline-as-code make it a popular choice among development teams. Its web-based user interfaces and distributed builds, make it easy to manage and monitor the build, test, and deployment process, which leads to more efficient and streamlined software development.

### IV. Proposed Framework
In this section, a proposed framework is presented that will make continuous development and continuous integration of software more traceable, secure, and trustworthy even though the teams are scattered in different locations. In this framework, the blockchain has been used make DevOps more secure, transparent, traceable and immutable without disturbing the automation of the processes of DevOps in any way. The main goal of this study is to bring data like source code, files, and artifacts to a single place so that it can become accessible easily while still being tamperproof and secure.

### A. *High-Level Architecture*

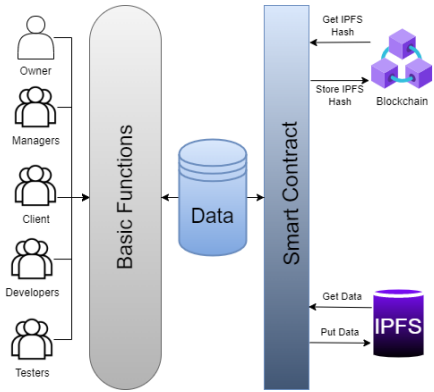

*Figure 1 High-Level Architecture*

The high-level architecture for the proposed framework is shown in Figure 1. It shows the overall working of the model that will use Blockchain, an InterPlanetary File System, and a Smart contract. In the architecture, the main focus is on data being decentralized without being in control of a central authority.

### B. *The Layered Architecture*

The suggested framework adheres to the blockchain architectural style, and we have presented the 6 layered architecture for distributed DevOps in Figure 2.

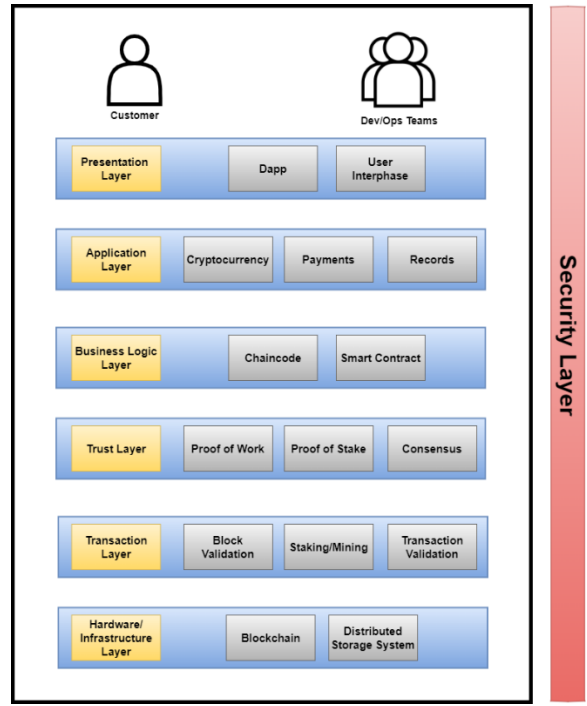

*Figure 2 Layered Architecture for Distributed DevOps*

1) PRESENTATION LAYER
The proposed system's Presentation layer includes a user-facing application and a decentralized application. Its primary goal is to connect clients and DevOps teams to the system.

2) APPLICATION LAYER
The metadata relating to transaction records, payments, mockups, prototypes, etc. along with agreements made between DevOps teams and clients in the form of videos, text, and audio, are present within this layer. Digital currencies, including ETH, USDT, BUSD, and BTC, are also in this layer to facilitate transactions after the completion of one successful iteration. The primary function of this layer is to enable seamless communication among stakeholders while serving as an intermediary between the Presentation Layer and Business Logic Layer.

3) BUSINESS LOGIC LAYER
This layer has all the Smart Contracts that govern the terms and conditions of interactions within the system. It serves as an active database for these contracts, enabling the acknowledgment, execution, and enforcement of communication rules. This layer plays a critical role in facilitating the functioning of the system by ensuring that all interactions are carried out in accordance with established rules and regulations.

4) TRUST LAYER

The Trust Layer of the layered architecture of distributed DevOps is responsible for managing the system's consensus algorithms such as Proof of Stake or Proof-of-Work. The trust layer plays a critical role in ensuring the security and reliability of the system by implementing robust consensus algorithms and security protocols.

5) TRANSACTION LAYER
The transaction layer is responsible for facilitating the developers and customers by enabling them to trigger transactional smart contracts. This layer also oversees the processes of mining/staking and validating the blocks containing these transactions. The transaction layer plays a critical role in the functioning of the proposed system.

6) HARDWARE/INFRASTRUCTURE LAYER
This layer includes the peer-to-peer network that validates transactions and a distributed storage system that stores and retrieves files on the decentralized web.

7) SECURITY LAYER
The Security Layer is responsible for security measures to protect the network from potential attacks. The security layer works in parallel with the rest of the system, incorporating algorithms and security protocols to safeguard the blockchain network.

### C. Layered Structure

The proposed framework consists of 6 layers, with each layer's transactions being recorded on the blockchain, thereby notifying all members of the blockchain network. The Layered Structure of system is shown in Figure 3.

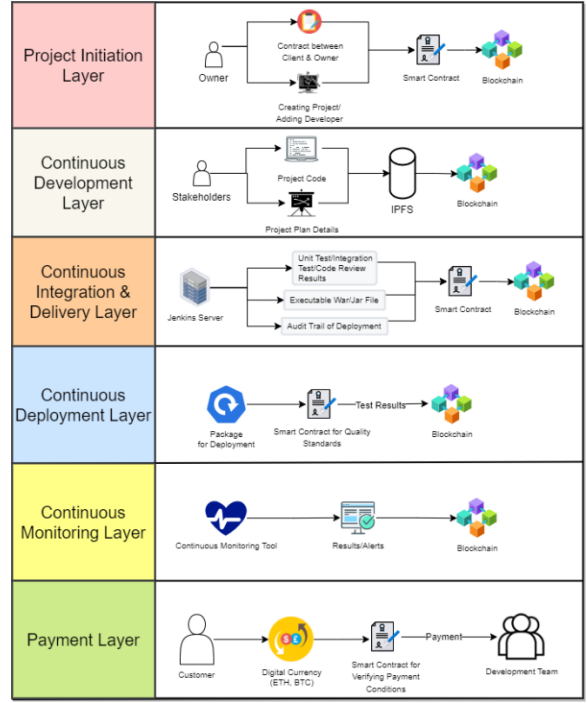

*Figure 3 Layered Structure*

### D. Detailed Architecture

This section describes the detailed structure of the planned framework. Every step of DevOps is described in detail with the involvement of Blockchain technology.

1) PROJECT INITIATION
The manager or owner can create the project at the project's initiation. After creating a project, the manager/owner can add members to the project using Client API. All the members will be given identity materials like key pairs generated by a Key Generator. Every developer will get a unique key, so there is no duplication problem. By using unique keys, developers can log in to the project to do their tasks. All unique keys and basic project details is going to upload on the blockchain using a smart contract.

All basic requirements and terms and conditions will be written at the initiation of the project by the owner or manager. All terms and conditions will store in the blockchain using a smart contract after the consensus is met between the manager/owner and client. Both customers and developers have to accept the terms and conditions set by the other side to reach a consensus. The complete work of the project initiation phase is shown in Figure 3.

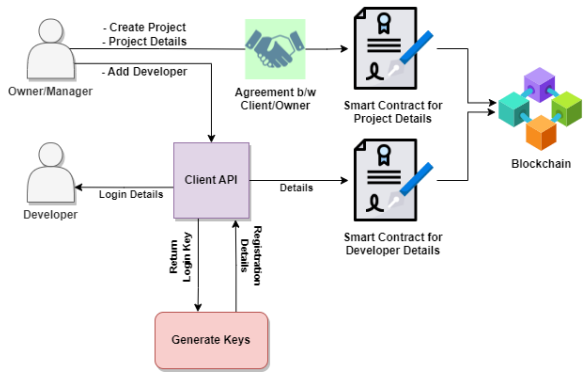

*Figure 4 Project Initiation Phase*

**2) CONTINUOUS DEVELOPMENT PHASE**

In the context of DevOps, the continuous development phase refers to the process of continuously planning and coding.

**Planning**: In this phase, the developer, managers, owner, and all stakeholders directly or indirectly involved in the planning and coding of software make a plan on how to complete the project successfully. Because the teams are located in different locations, the planning can happen via online video calling platforms like Zoom or Google Meetings. All the details of the planning phase, like video conference recordings and notes can be stored in the IPFS because of Blockchain's scalability issue. Only the returned hash file is stored on the Blockchain to make it tamperproof and more secure so that no one can change it, as shown in Figure 4. This data is visible to all the stakeholders. In this way, anyone who can access details can see them but not change them.

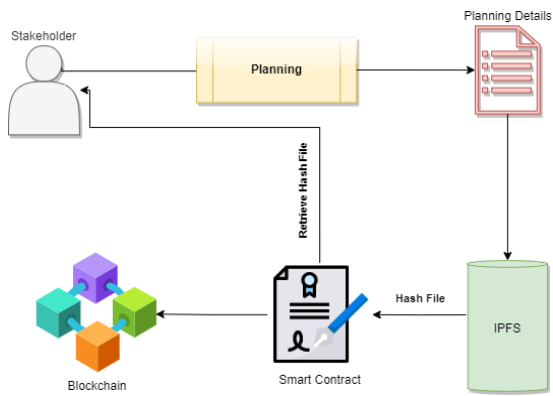

*Figure 5 Planning for Continuous Development*

**Coding**: Git has been one of the most popular tools for decentralized source code management and is widely used in DevOps. However, the InterPlanetary File System (IPFS) can be used to do the enhance decentralization of Git. IPFS provides a content-addressed version control system for managing files. It is possible to Serve a Git repository through the IPFS network to provides a decentralized and distributed way to host and share Git repositories and make changes to it.

**3) CONTINUOUS INTEGRATION & DELIVERY PHASE**

Continuous Integration (CI) and Continuous Delivery (CD) are essential components of a DevOps workflow, enabling teams to build, test, and deploy software quickly and reliably. In this blockchain-based framework, CI/CD can be further enhanced by leveraging the immutability and transparency of the blockchain.

To implement CI/CD developers can use open-source tools such as Jenkins. Jenkins is one of the most used tools for developers [14]. Jenkins is an automation server that can be used to automate parts of the software development process, such as building, testing, and releasing code changes. Whenever developers update the repository, the code goes through the Jenkins server, which performs a series of automated operations, including code review, unit testing, integration testing, and building an executable package (e.g., WAR or JAR).

Once the package is built, it is uploaded to a decentralized storage system IPFS. The IPFS hash of the package, along with the results of the code review, unit tests, and integration tests, can be stored on the blockchain. This provides a transparent and immutable record of the code changes and the testing that has been performed. If there is a code error during code review, unit test, or integration testing, the system generates an alert that notifies all developers to remove the error in the code and update the repo as soon as possible.

To enable continuous delivery, smart contracts can be used. A smart contract can be created that automatically deploys the package to a specified server or cloud provider when certain conditions are met (e.g., passing all tests). The smart contract can also track the deployment process, ensuring that the package is delivered correctly and providing an audit trail of the delivery.

The following information of executable file that can be stored on blockchain are shown in Table 2

| Info | Description |

| Time | The time when files upload |
|---|---|
| Date | Date of uploading of file |
| Name | Name of file |
| Version | Version of file |

After the file is uploaded on Blockchain, all developers get notified about the file. The complete working of CI/CD is shown in Figure 5.

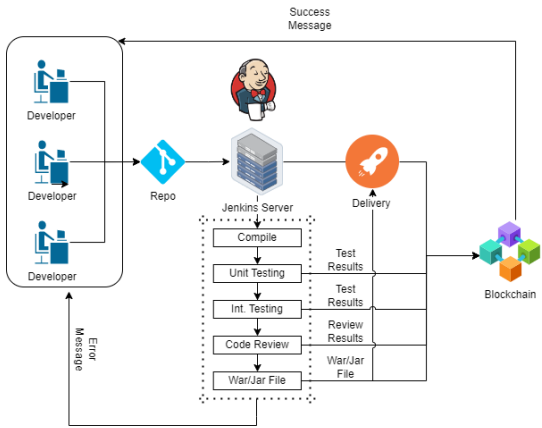

*Figure 6 CONTINUOUS INTEGRATION & DELIVERY PHASE*

### 4) CONTINUOUS DEPLOYMENT

Continuous Deployment is a software development strategy that involves the automatic release of code changes to the production environment. Smart Contracts play a major role in implementing CD in our proposed framework.

Before deploying the main file, the code has to go through a smart contract specifically created to check quality standards, security requirements, and compliance with regulatory requirements. The smart contract should be already integrated into the CI/CD pipeline to enforce the deployment criteria specified in the contract. This can be done by creating a custom script or plugin that interacts with the contract or by using an existing blockchain integration tool. Once the main file is deployed, the owners, managers, clients, and developers are notified about this change. The status of this implemented functionality changes to "Deployed".

### 5) CONTINUOUS MONITORING

In this phase, all the metrics related to the performance and availability of the software and infrastructure components can be recorded on the blockchain with the help of smart contracts. By recording these metrics on the blockchain, they become immutable and transparent.

When a problem or alert is detected in the system, first it is recorded on the blockchain, and then all the developers who are part of the DevOps team are notified about it immediately. This notification ensures that the problem can be stored and then resolved as soon as possible, minimizing the impact on end-users.

To monitor the process, any preferred software can be used. However, the software must be compatible with the blockchain framework being used in the distributed DevOps environment. This can be done by creating a custom script or plugin. This ensures that the monitoring data can be recorded on the blockchain and accessed by all the relevant stakeholders securely and transparently.

### 6) PAYMENTS PHASE

The payment phase automatically gets enabled when one iteration gets completed by the organization or after two weeks. The payment can be set to be paid after 2 weeks or after one successful deployment of functionality, depending on the decision made at the time of project initiation.

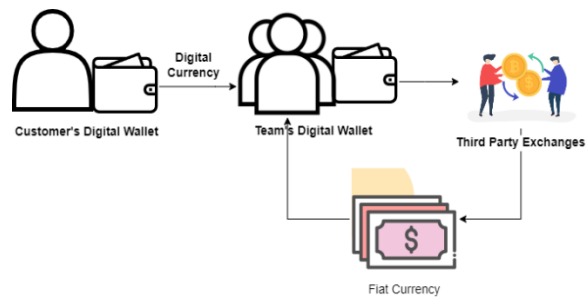

*Figure 6 PAYMENTS PHASE*

To complete the payment process, every stakeholder in the project must have a Digital Wallet. The payment can be made with digital currencies like ETH (Ethereum) or BTC (Bitcoin). The managers, developers, or testers can change this digital currency to their local currency, like PKR, USD, or EUR, using cryptocurrency exchanges, as shown in Figure 6.

The client has to pay the payment; otherwise, as a result, the work cannot proceed further. The client must pay the team within the time frame that was settled while initiating the project. The payment requirements are written in a smart contract, so they cannot be changed or altered. The JSON object for payment in Table 1 is shown.

**Agreement**
{
   "Project Budget": "$1000"
   "Payment After 1 Iteration": "$100"

OR
"Payment After 2 Weeks": "$100"
"In Case of Non Payment": "Stop Project's Functions"
}

*Table 1. JASON File for Agreement*

## V. Implementation and Performance

In this section, we have tested the efficiency of the framework to demonstrate its effectiveness in a real-world scenario.

### A. Performance Assessment

For measuring benchmarks, the Hyperledger Caliper is used which is a blockchain benchmarking tool. For implementing the proposed framework for demonstrational purposes, the Hyperledger Fabric has been used and the test was run using an AWS EC2 instance. Hyperledger Fabric Network consists of 4 organizations, and 1 orderer, and the stated database used in the HLF network is CouchDB, a document-oriented database that provides robust storage capabilities.

### B. Performance Evaluation

In this section, the benchmarks are presented to evaluate the expected efficiency of the stated model. The efficiency is measured and evaluated utilizing three crucial performance indicators: TPS (transactions per second), Throughput (number of successful transactions per second), and Latency (time taken per transaction to receive a response). The results are meticulously presented in Table 2 to provide a comprehensive understanding of the performance.

Caliper report

| Name | Succ | Fail | Send Rate (TPS) | Max Latency (s) | Min Latency (s) | Avg Latency (s) | Throughput (TPS) |
|---|---|---|---|---|---|---|---|
| Round0-QueryPrivateData-TxNumber-FixedRate | 2500 | 0 | 143.7 | 0.30 | 0.01 | 0.02 | 143.6 |
| Round1-QueryPrivateData-TxDuration-FixedRate | 4801 | 0 | 80.0 | 0.07 | 0.00 | 0.01 | 80.0 |
| Round2-QueryPrivateData-TxNumber-Linearate | 2500 | 0 | 135.3 | 0.31 | 0.01 | 0.03 | 135.2 |
| Round3-QueryPrivateData-TxDuration-Linearate | 3988 | 0 | 66.5 | 0.04 | 0.00 | 0.01 | 66.5 |

*Table 2. Caliper Report*

The results of these benchmarks revealed that the latency of transactions remained low and consistent, regardless of the combination of tests used. This demonstrates the robustness of the system and its ability to process transactions quickly. Additionally, the transactions per second and overall throughput also remained stable, indicating that the system is able to handle a high volume of transactions without any slowdown.

Furthermore, the resource utilization for every round is also shown in Table 1, 2, 3, and 4.

| Name | CPU% (max) | CPU% (avg) | Memory(max) [MB] | Memory(avg) [MB] | Traffic In [MB] | Traffic Out [MB] | Disc Write [B] | Disc Read [B] |
|---|---|---|---|---|---|---|---|---|
| /peer0.org4.example.com | 1.46 | 1.16 | 42.4 | 42.4 | 0.0500 | 0.0464 | 0.00 | 0.00 |
| /peer0.org3.example.com | 1.3 | 1.18 | 42.0 | 42.0 | 0.0477 | 0.0432 | 0.00 | 0.00 |
| /peer0.org2.example.com | 1.96 | 1.65 | 63.8 | 63.8 | 0.0678 | 0.0718 | 0.00 | 0.00 |
| /peer0.org1.example.com | 34.97 | 18.73 | 60.1 | 60.1 | 6.03 | 9.53 | 0.00 | 0.00 |
| /orderer.example.com | 0.08 | 0.07 | 9.50 | 9.50 | 0.00320 | 0.00344 | 0.00 | 0.00 |

*Table 3. Resource Utilization for Round 0*

| Name | CPU% (max) | CPU% (avg) | Memory(max) [MB] | Memory(avg) [MB] | Traffic In [MB] | Traffic Out [MB] | Disc Write [B] | Disc Read [B] |
|---|---|---|---|---|---|---|---|---|
| /peer0.org4.example.com | 1.42 | 0.94 | 42.4 | 42.4 | 0.139 | 0.132 | 0.00 | 0.00 |
| /peer0.org3.example.com | 2.1 | 1.12 | 42.0 | 42.0 | 0.142 | 0.134 | 0.00 | 0.00 |
| /peer0.org2.example.com | 1.71 | 1.09 | 63.8 | 63.8 | 0.193 | 0.204 | 0.00 | 0.00 |
| /peer0.org1.example.com | 22.32 | 17 | 60.1 | 60.1 | 12.5 | 18.8 | 0.00 | 0.00 |
| /orderer.example.com | 0.13 | 0.07 | 9.50 | 9.50 | 0.00595 | 0.00580 | 0.00 | 0.00 |

*Table 4 Resource Utilization for Round 1*

| Name | CPU% (max) | CPU% (avg) | Memory(max) [MB] | Memory(avg) [MB] | Traffic In [MB] | Traffic Out [MB] | Disc Write [B] | Disc Read [B] |
|---|---|---|---|---|---|---|---|---|
| /peer0.org4.example.com | 1.32 | 1.1 | 42.4 | 42.4 | 0.0478 | 0.0447 | 0.00 | 0.00 |
| /peer0.org3.example.com | 1.43 | 1.25 | 42.0 | 42.0 | 0.0486 | 0.0464 | 0.00 | 0.00 |
| /peer0.org2.example.com | 1.74 | 1.4 | 63.8 | 63.8 | 0.0678 | 0.0703 | 0.00 | 0.00 |
| /peer0.org1.example.com | 40.6 | 22.9 | 60.1 | 60.1 | 6.06 | 9.42 | 0.00 | 0.00 |
| /orderer.example.com | 0.1 | 0.07 | 9.50 | 9.50 | 0.00320 | 0.00343 | 0.00 | 0.00 |

*Table 5 Resource Utilization for Round 2*

| Name | CPU% (max) | CPU% (avg) | Memory(max) [MB] | Memory(avg) [MB] | Traffic In [MB] | Traffic Out [MB] | Disc Write [B] | Disc Read [B] |
|---|---|---|---|---|---|---|---|---|
| /peer0.org4.example.com | 1.23 | 1.03 | 42.4 | 42.4 | 0.138 | 0.130 | 0.00 | 0.00 |
| /peer0.org3.example.com | 1.46 | 1.14 | 42.0 | 42.0 | 0.142 | 0.135 | 0.00 | 0.00 |
| /peer0.org2.example.com | 2.92 | 1.63 | 63.8 | 63.8 | 0.200 | 0.208 | 0.00 | 0.00 |
| /peer0.org1.example.com | 19.95 | 15.03 | 60.1 | 60.1 | 10.4 | 15.7 | 0.00 | 0.00 |
| /orderer.example.com | 0.13 | 0.08 | 9.50 | 9.50 | 0.00595 | 0.00580 | 0.00 | 0.00 |

*Table 6 Resource Utilization for Round 3*

The results of the resource utilization analysis reveal a consistent pattern across all four rounds. It is noteworthy that in each round, the utilization of CPU, RAM, and Memory by Organization 1 was consistently higher than the other organizations, indicating a higher level of resource utilization by the first organization.

However, it is important to note that as the number of transactions per second increases, variations in latency, throughput, or resource utilization may occur.

## VI. Discussion

In this section, the performance results are formally discussed. This framework is useful in successfully making software projects using DevOps. The proposed framework eliminates the current problems of transparency, trackability, security, communication, and collaboration present in Distributed DevOps.

The various benefits of this model are;

### A. Enhanced security
Blockchain technology can help to enhance the security of software development processes by providing a tamper-evident record of code changes and deployments. This can help to ensure that only authorized code changes are deployed and that the code changes have not been tampered with or modified in any way.

### B. Improved transparency
This framework can provide a transparent record of code changes and deployments, making it easier for teams to track and audit the software development process. This can help to improve trust and collaboration within the development team, and can also help to meet regulatory and compliance requirements.

### C. Enhanced traceability
It can help to improve the traceability of code changes and deployments, making it easier to track the origin and evolution of software projects. This can be particularly useful for organizations that need to meet strict traceability requirements, such as in the aerospace, defense, and healthcare industries.

### D. Improved collaboration
The collaboration within development teams can be improved by providing a shared, tamper-evident record of code changes and deployments. This can help to reduce the risk of conflicts and errors and can make it easier for teams to work together on complex projects.

### E. Increased efficiency
By automating the tracking and verification of code changes and deployments using blockchain technology, organizations can streamline their software development processes and increase efficiency. This can help to reduce the time and effort required to release new features and updates, and can also help to improve the reliability and quality of software projects.

There is no doubt that there are some limitations involved in this model. The processes can get a lot easier but the performance and speed of uploading or loading data can be slow because Blockchain is relatively slower as compared to traditional databases. The implementation cost for this type of framework is another drawback. Organizations working on a limited budget cannot implement this framework for software development due to some budgeting constraints and the complexity of the model.

Overall, the evaluations of this research disclose that a major technological advance known as Blockchain has great potential to change how we used to do software development.

## VII. Conclusion and Future Work

The current distributed DevOps for software development is lacking in terms of security, data protection, privacy, transparency, and trackability. Because of these shortcomings, the operations and development teams cannot work together properly. That is why this model has been proposed that will allow the collaboration or development and operation teams more smoothly and problem free. The framework utilizes smart contracts and a decentralized architecture to enable secure and efficient collaboration among distributed teams in the development and operations of software systems. We discussed the benefits of using blockchain technology in DevOps, such as increased transparency, enhanced traceability, improved collaboration, and increased efficiency. We also presented an implementation and performance of the framework by using Hyperledger Fabric that demonstrated the effectiveness of the proposed framework in a real-world scenario.

There are several directions for future work that can be pursued based on the proposed framework. One potential area of research is to further investigate the framework with other emerging technologies, such as artificial intelligence and machine learning, to enhance its capabilities and automation. Additionally, it will be interesting to study the real-world adoption and usage of the framework by different organizations and industries. Furthermore, there is also a scope to explore the regulatory compliance aspect of the framework and its adoption in different geographical regions. Overall, the proposed framework provides a promising solution for addressing the challenges of collaboration in modern software development and operations, and there is much potential for further research and development in this area.